\documentclass{article}
\usepackage{arxiv}
\usepackage[utf8]{inputenc} 
\usepackage[T1]{fontenc}    
\usepackage{hyperref}       
\usepackage{url}            
\usepackage{booktabs}       
\usepackage{amsfonts}       
\usepackage{nicefrac}       
\usepackage{microtype}      
\usepackage{lipsum}
\usepackage{graphicx}
\graphicspath{ {./images/} }
\usepackage{algorithm}
\usepackage{algpseudocode}
\usepackage{amsmath}

\title{Forecasting Commodity Prices Using
Long Short-Term Memory Neural Networks
}

\author{
 Racine Ly \\
  AKADEMIYA2063\\
  Kigali, Rwanda \\
  \texttt{rly@akademiya2063.org} \\
   \And
 Fousseini Traore \\
  International Food Policy Research Institute \\
  Dakar, Senegal \\
  \texttt{fousseini.traore@cgiar.org} \\
  \And
 Khadim Dia \\
  AKADEMIYA2063\\
  Kigali, Rwanda \\
  \texttt{kdia@akademiya2063.org}\\
}

\makeatletter
\renewcommand\@biblabel[1]{}
\renewenvironment{thebibliography}[1]
     {\section*{\refname}%
      \@mkboth{\MakeUppercase\refname}{\MakeUppercase\refname}%
      \list{}%
           {\leftmargin0pt
            \@openbib@code
            \usecounter{enumiv}}%
      \sloppy
      \clubpenalty4000
      \@clubpenalty \clubpenalty
      \widowpenalty4000%
      \sfcode`\.\@m}
     {\def\@noitemerr
       {\@latex@warning{Empty `thebibliography' environment}}%
      \endlist}
\makeatother

\begin{document}
\maketitle
\begin{abstract}
This paper applies a recurrent neural network (RNN) method to forecast cotton and oil prices. We show how these new tools from machine learning, particularly Long-Short Term Memory (LSTM) models, complement traditional methods. Our results show that machine learning methods fit reasonably well the data but do not outperform systematically classical methods such as Autoregressive Integrated Moving Average (ARIMA) models in terms of out of sample forecasts. However, averaging the forecasts from the two type of models provide better results compared to either method. Compared to the ARIMA and the LSTM, the Root Mean Squared Error (RMSE) of the average forecast was 0.21 and 21.49 percent lower respectively for Cotton. For Oil the forecast averaging does not provide improvements in terms of RMSE. We suggest using a forecast averaging method and extending our analysis to a wide range of commodity prices. 
\end{abstract}


\section{Introduction}
Forecasting commodity prices is paramount for many economic actors. When building budgets, experts rely on growth projections at the government level, which are almost always based on underlying forecasts of primary commodities exported by the country. Oil-dependent countries represent a typical example where this kind of scenario is encountered. For developing countries, many depend on a few raw materials (agricultural and minerals), the price of which determines the growth rate. Being able to forecast commodity prices is vital for other public entities or parastatals too. In many developing countries, parastatals are managing stabilization funds aimed at smoothing commodity price movements. These entities need world price forecasts in order to fix producers' prices for the current campaign. For researchers, knowing the best data generating process and the forecast errors is essential for various modeling purposes. For instance, in agricultural models involving expectations \footnote{The widely used Nerlove model falls in that category.} , the expected price is generally given by an ARMA model (Antonovitz and Green, 1990; De Janvry and Sadoulet, 1995). As we will show later, LSTM networks provide a good alternative to ARMA models. \\
\newline
Since the seventies, Box-Jenkins (1970)\footnote{Although one can mention earlier work by Tinbergen (1939) and Klein (1950).} approaches have been popular in forecasting time series. These approaches have introduced ARMA models and their extensions as the cornerstone of forecasting tools. However, machine learning methods that can handle time-series data and perform forecasting have grown over the last three decades. Among all methods, those based on Recurrent Neural Network are particularly interesting as they can carry over information (memory) from previous periods into the future. Early papers using RNNs to forecast time series include Kamijo and Tanigawa (1990), Chakraborty et al. (1992), and a comparison with ARIMA models by Kohzadi et al. (1996).  However, one issue encountered with the first generation of RNN is the so-called vanishing gradient problem for highly dependent (long memory) data. Thus, in a seminal paper, Hochreiter and Schmidhuber (1997) proposed a new approach called Long Short-term Memory with the capacity to filter out which information from the past should be processed and retained\footnote{See section 2 for a full description of LSTM networks.} . This triggered new literature on forecasting times series (Gheyas and Smith 2011; Khandelwal et al., 2015; Kumar et al., 2018). In a similar way, other classes of machine learning methods such as Support Vector Machine (SVM) regressions, have been developed and applied to time series forecasting, including hybrid approaches with ARIMA models (Pai and Lin, 2005).  \\
\newline
Our objective in this paper is to focus on LSTM models as they present a series of interesting characteristics. First, they are non-parametric so that they can handle suitably non-linear patterns. Second, they do not require the error term to follow a distribution. Third, LSTM models do not require the underlying data to follow a stationary process, so they are not affected by unit-roots. These three features present an interesting advantage over regression-based methods, be they ARIMA or not. Thus, using monthly data of cotton and oil prices, our results suggest that the LSTM model fits the data reasonably well but does not outperform the ARIMA models for out of sample forecast. However, a combination of forecasts from the two models yields better results for the cotton dataset than either approach. \\
\newline
The remainder of this paper is organized as follows. We first present in section 2 the traditional Recurrent Neural Networks (RNN), and then, the LSTM approach proposed to solve the vanishing gradient problem of the first generations of RNN. In section 3, we present the data set used. Section 4 presents the computational aspects while the results of the LSTM approach applied to oil and cotton prices are highlighted in section 5. In the sixth section, we compare the LSTM approach with ARIMA models. The seventh section the forecast averaging approach and our conclusions are discussed in section 8. 

\section{Methodology}
\subsection{Artificial Neural Networks}
An artificial Neural Network is a supervised learning technique within the machine learning set of models. It is used for both regression and classification tasks depending on the type of data used and the modeling purpose. An artificial neural network comprises an input layer that encounters the different inputs' features, at least one hidden layer that will process the dataset's hidden characteristics, and an output layer that will yield the network's forecasts. Each layer has several units called "neurons," which have the role of receiving information from preceding neurons and send processed information to the following ones. The input layer has several neurons equal to the number of the dataset's features, and each layer is augmented with a so-called "bias" neuron which role is to facilitate the algorithmic writing. The number of hidden layers – which is one of the so-called hyper-parameters - is set by the modeler through a fine-tuning process. The number of output layer's neurons is equal to the number of expected outputs from the network.\\
\newline
Within an artificial neural network, each neuron communicates with all the neurons from the preceding layer and with those from the following layer (aside from the last layer) through weights affected at each connection. The latter is computed by applying an activation function - which is usually a sigmoid or hyperbolic tangent function, among others that are used in the literature - to the product of a weights vector controlling the mapping function from a layer $l$ to the following layer $l+1$, and the preceding activations of the previous layer.\\
\newline
The learning process of an artificial neural network is performed using the backpropagation algorithm (Rumelhart, Hinton, and William, 1986), which is the optimization process through which the weights that represent the connections between neurons are updated using the chain rule method to compute the gradient of the loss function with respects to the weights. A gradient descent rule or related version is used to update their values. The weights are randomly initialized using a uniform distribution, the bias units are set to zero, and the initial activations for the first layer are equal to the dataset's features.\\
\newline
The backpropagation procedure is as follows: A forward pass is performed by computing the "cash" values used as an argument to assess the activation of the layers. For an $L$-layers neural network, the $k^{th}$ layer uses the following constitutive equations to yield the cash vector  $z^{\left[k\right]}$ and the activation vector $a^{\left[k\right]}$ for each neuron and layer.

\begin{eqnarray}
z^{\left[k\right]}&=&\eta^{\left[k\right]}a^{\left[k-1\right]}+b^{\left[k\right]} \label{eq1} \\
a^{\left[k\right]}&=&g^{\left[k\right]}\left(z^{\left[k\right]}\right) \label{eq2}
\end{eqnarray}

In equations \eqref{eq1} and \eqref{eq2}, $\eta^{\left[k\right]}$ is the vector of weights of all connections going from neurons at layer $k-1$ to neurons at layer $k$, and $g^{\left[k\right]}$ is the activation function that is applied to neurons at layer $k$. The network's last activation function corresponds to the network's predicted value, which is compared with the actual value from the response variable provided by the training dataset. A backward pass is then applied to the artificial neural network to compute the gradients' cost function for weights assigned to each connection. The residual $\delta[L] = a[L] - y$, at the last layer between the network predictions $a[L]$ and actual values $y$ is backpropagated into the network, and the residual associated with the preceding layers is computed as follows:

\begin{eqnarray}
\delta^{\left[k\right]}&=&\left(\eta^{\left[k\right]}\right)^T\delta^{\left[k+1\right]}\circ{g^{\left[k\right]}}^\prime\left(z^{\left[k\right]}\right) \label{eq3}
\end{eqnarray}

The gradient of the cost function $\mathbf{J}$ for the weight between two nodes $i$ and $j$ at a layer $k$ materialized by $\eta_{ij}^{[k]}$, is equal to the product of the activation of the node $j$ at layer $k$ and the residual of the node $i$ at layer $k+1$ computed with equation \eqref{eq3}.

\begin{eqnarray}
\frac{\partial \mathbf{J}(\eta)}{\partial \eta_{ij}^{[k]}} &=& a_{j}^{[k]} \delta_{i}^{[k+1]} \label{eq4}
\end{eqnarray}

For an $L$-layers artificial neural network with m examples in the training set ${(x^{(1)},\ y^{(1)}),\ ...\ ,\ (x^{(m)},\ y^{(m)})}$, an iterative process is performed to aggregate the gradients in an accumulator using equation \eqref{eq4}. The procedure is as follows:

\begin{algorithm}
  \caption{Backpropagation Algorithm}\label{BackProp}
  \begin{algorithmic}[c]
  \For{\texttt{$i = 1$ to $m$}}
        \State \texttt{Set $a^{\left[1\right]}=x^{\left[i\right]}$}
        \State \texttt{Forward propagation to compute $a[k]$ for $k = 1, 2, 3, \cdots, L$}
        \State \texttt{Compute $\delta[L] = a[L] - y[i]$}
        \State \texttt{Compute $\delta[L-1],\cdots,\delta[2]$ with equation \eqref{eq3}}
        \State \texttt{$\Delta_{ij}^{[k]}$:=$\Delta_{ij}^{[k]} + a_{j}^{[k]} \delta_{i}^{[k+1]}$}
        \State \texttt{$D_{ij}^{[k]} = \frac{1}{m} \left(\Delta_{ij}^{[k]} + \lambda \eta_{ij}^{[k]}\right)$ if  $j \neq 0$ (nonbias node)}
        \State \texttt{$D_{ij}^{[k]} = \frac{1}{m} \left(\Delta_{ij}^{[k]}\right)$ if  $j = 0$ (for bias node)}
        \State \texttt{$\frac{\partial \mathbf{J}(\eta)}{\partial \eta_{ij}^{[k]}} = D_{ij}^{[k]}$}
      \EndFor
  \end{algorithmic}
\end{algorithm}

Artificial neural networks are suitable for numerical and qualitative datasets for regression and classification tasks. However, for sequential-type datasets where inputs and outputs are sequences, ANNs are not adequate for two main reasons: (i) Inputs and outputs need to have the same lengths and such could not be the case in sequential datasets, (ii) ANNs cannot deal with the need of sharing features learned across different positions in a sequential dataset. Recurrent Neural Networks (RNN) have been introduced to solve the limitations mentioned above.

\subsection{Recurrent Neural Networks and Gated Recurrent Units (GRU)}

The primary purpose of Recurrent Neural Networks is to deal with datasets that have inputs and outputs that are sequences. RNNs are built on top of the same ideas and architecture of an artificial neural network. The main difference resides in RNNs being able to encounter historical and current sequences in predicting the outcome at the same current time-step or sequence. Such confer to RNNs a real advantage in predicting time-sensitive sequential data such as time-series.\\
\newline
For a sequential labeled dataset $(X,Y)$, where $X \in \mathbb{R}^{T_{x}}$ and $Y \in \mathbb{R}^{T_{y}}$, where $T_{x}$ and $T_{y}$ are the number of sequences in the input and labels data respectively, the guiding equations to perform a forward pass within RNNs is as follows:

\begin{eqnarray}
a^{<t>} &=& g(W_{aa}a^{<t-1>} + W_{ax}x^{<t>} + b_{a}) \label{eq5} \\
y^{<t>} &=& g(W_{ya}a^{<t>} + b_{y}) \label{eq6}
\end{eqnarray}

Where $a^{<t>}$ is the activation value for the RNN unit at the sequence $<t>$, $W_{aa}$, $W_{ax}$ and $W_{ya}$ are a set of weights that materialized the relationship between activations from one sequence to the previous one; activations and inputs sequences; and outputs and activations from the same sequence, respectively. The parameters $b_{a}$ and $b_{y}$ are the bias terms.\\
\newline
A significant limitation of RNNs resides in their insufficient capacity to encounter long-term dependencies. Such usually appears with a problem of vanishing gradients. The latter occurs when the error between the network outputs and the actual data is poorly backpropagated through the network and tends not to affect the first layers for relatively deep architecture. Therefore, the low gradient values make the network's weights update not occurring, which yields inaccurate predictions. The opposite, exploding gradient, can also be a limitation of RNNs in dealing with long-term dependencies. However, such could be mitigated using the gradient clipping technique, which consists of imposing a maximum limit to gradient values and avoiding numerical overflows in weights' values.\\
\newline
To address the vanishing gradient issue of RNNs, Gated Recurrent Units (GRUs) have been proposed by (Cho, Van Merriënboer, Bahdanau, and Bengio, 2014) and (Chung, Gucehre, Cho, and Bengio, 2014). The main improvement compared to RNNs is introducing a memory cell that will drag, through the network, long-term dependencies from previous sequences and mitigate the vanishing gradient issue. For such, a memory cell at a sequence $<t>$, noted $c^{<t>}$ takes the role of the activation values. At each sequence, a candidate for the memory cell is computed using a tangent hyperbolic activation function applied to a linear combination of the input sequence $x^{<t>}$ and the previous memory cell value $c^{<t-1>}$. A gate $\Gamma_{u}$ is computed to assess if the memory cell's information at the previous sequence $c^{<t-1>}$ will be kept or replaced by the memory cell candidate. The gate is computed by applying a sigmoid activation function to the linear combination of $c^{<t-1>}$ and $x^{<t>}$. Hence, the new memory cell at the sequence $<t>$ is computed using the mixture rule. The guiding equations of a GRU in its simplest form are as follows:

\begin{eqnarray}
\tilde{c}^{<t>} &=& tanh(W_{c}[c^{<t-1>},x^{<t>}] + b_{c}) \label{eq7} \\
\Gamma_{u} &=& \sigma(W_{u}\left[c^{<t-1>},x^{<t>}\right] + b_{u}) \label{eq8} \\
c^{<t>} &=& \Gamma_{u} \circ \tilde{c}^{<t>} + (1 - \Gamma_{u}) \circ c^{<t-1>} \label{eq9}
\end{eqnarray}

The main advantage of GRU in mitigating the vanishing gradients issue with RNNs resides in the factor $(1- \Gamma_{u})$ in which even if $\Gamma_{u}$ is very close to zero, the value of $c^{<t>}$ in equation \eqref{eq9} is maintained, which will avoid the gradient of the cost function for activations (cf. equation \eqref{eq4}) to be null, therefore, allowing the updates of the weights using gradient descent-based rule.\\
\newline
Equations \eqref{eq7} to \eqref{eq9} are the constitutive ones for the simplest version of the GRU. A complete version consists of adding a gate to assess if the memory cell at the previous sequence $c^{<t-1>}$ is still relevant for each unit at the sequence $<t>$. The new set of equations becomes, with all weights and biased being updated using the backpropagation algorithm:

\begin{eqnarray}
\tilde{c}^{<t>} &=& tanh(W_{c}[\Gamma_{r}\circ [c^{<t-1>},x^{<t>}] + b_{c}) \label{eq10} \\
\Gamma_{u} &=& \sigma(W_{u}[c^{<t-1>},x^{<t>}] + b_{u}) \label{eq11} \\
\Gamma_{r} &=& \sigma(W_{r}[c^{<t-1>},x^{<t>}] + b_{r}) \label{eq12} \\
c^{<t>} &=& \Gamma_{u} \circ \tilde{c}^{<t>} + (1-\Gamma_{u}) \circ c^{<t-1>} \label{eq13}
\end{eqnarray}

\subsection{Long-Short Term Memory (LSTM)}

The Gated Recurrent Unit is well suited to learn long connections in a sequence. Another version of the sequence model built on top of GRU's is the so-called Long Short-Term Memory model (LSTM) introduced by (Hochreiter and Schmidhuber, 1997). LSTM models introduce two new gates that make the learning process more flexible: the forget and output gates. For the first one, in LSTM modeling, instead of assigning the update gate's complement $1-\Gamma_{u}$ to control the memory cell at the previous sequence $c^{<t-1>}$ (cf. equation \eqref{eq9}), a new "forget" gate $\Gamma_{f}$ is introduced. The second "output" gate $\Gamma_{o}$ is to yield the content of the new memory cell ct. Hence, the set of equations that define the computations within an LSTM unit is:

\begin{eqnarray}
\tilde{c}^{<t>} &=& tanh(W_{c}[a^{<t-1>},x^{<t>}] + b_{c}) \label{eq14}\\
\Gamma_{u} &=& \sigma(W_{u}[a^{<t-1>},x^{<t>}] + b_{u}) \label{eq15} \\
\Gamma_{f} &=& \sigma(W_{f}[a^{<t-1>},x^{<t>}] + b_{f}) \label{eq16} \\
\Gamma_{o} &=& \sigma(W_{o}[a^{<t-1>},x^{<t>}] + b_{o}) \label{eq17} \\
c^{<t>} &=& \Gamma_{u} \circ \tilde{c}^{<t>} + \Gamma_{f} \circ c^{<t-1>}  \label{eq18} \\
a^{<t>} &=& \Gamma_{o} \circ tanh(c^{<t>}) \label{eq19}
\end{eqnarray}

Sequence modeling through a recurrent neural network and its most advanced version, LSTM, have the advantage over simple neural network architecture by its capacity to encounter current and previous sequences into producing forecasts. Moreover, as showcased by the constitutive equations illustrated above, LSTM modeling does not need the input dataset to be stationary, as required by the ARIMA-type models. Such makes LSTM a potential candidate in time-series forecasts with the possibility of encountering historical shocks across units and layers through the memory cell at each sequence, which is the time steps in time-series context.

\begin{figure}[htbp!]
  \caption{(a) A one-layer LSTM network for sequence modeling. (b) Illustration of computations within an LSTM unit at a time-step or sequence $<t>$}
  \centering
    \includegraphics[width=0.8\textwidth]{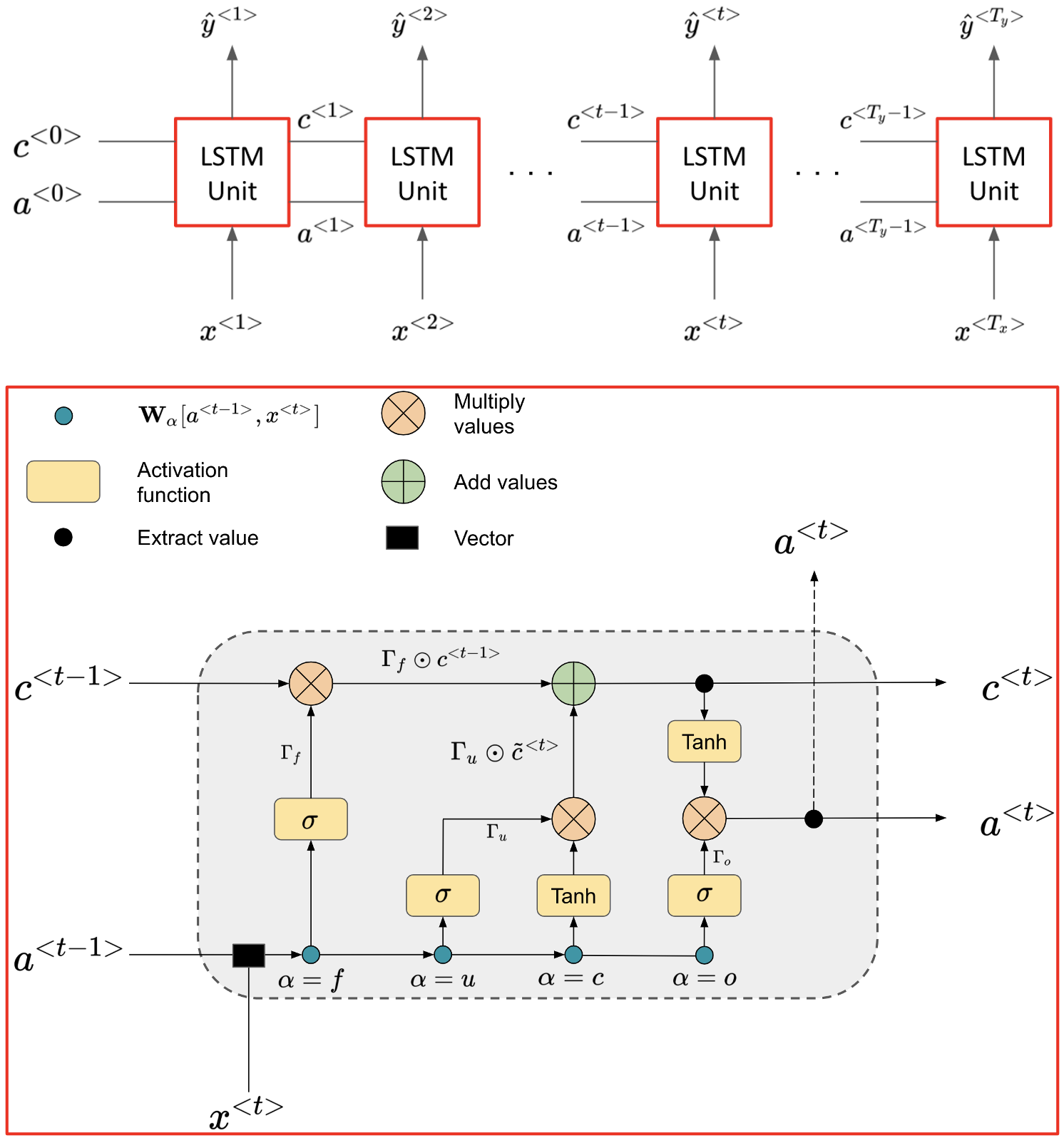}
\end{figure}

\section{Data} 

We use international prices for two commodities: one agricultural product (cotton) and one mineral (oil). All the data are taken from the World Bank commodity prices dataset. The Cotton price used is the Cotton A index, the most common index used to represent offering prices on the international cotton market. It corresponds to the average of the cheapest five quotations among eighteen varieties of upland cotton traded internationally. The base quality of the index is MIDDLING $1-1/8''$. For crude oil, we use the average spot price of Brent, Dubai, and West Texas Intermediate, the three leading indices representing the market. All prices are observed monthly. The cotton and oil datasets have 708 data points each, acquired from January 1960 to December 2018.

\subsection{Training and test datasets construction for LSTM modeling}

Sequential modeling for time-series forecasts using the LSTM framework requires transforming the dataset into features and labels for a supervised learning process. For such, the sliding window technique has been used across the cotton and oil datasets. It consists of shifting a window with a fixed-width – also called the lookback parameter - from left to right in the dataset. The window moves one unit at a time, and the data points within its range are taken as examples in the input features. The next value at the right edge of the window is considered as the label at each step; in other words, it is considered as the expected output produced by the data within the sliding window.

\begin{figure}[htbp!]
  \caption{Illustration of the sliding window technique to create input features and labels from time-series datasets. The dataset is read from left to right, and each increment in the window movement is considered an example of the training process}
  \centering
    \includegraphics[width=1\textwidth]{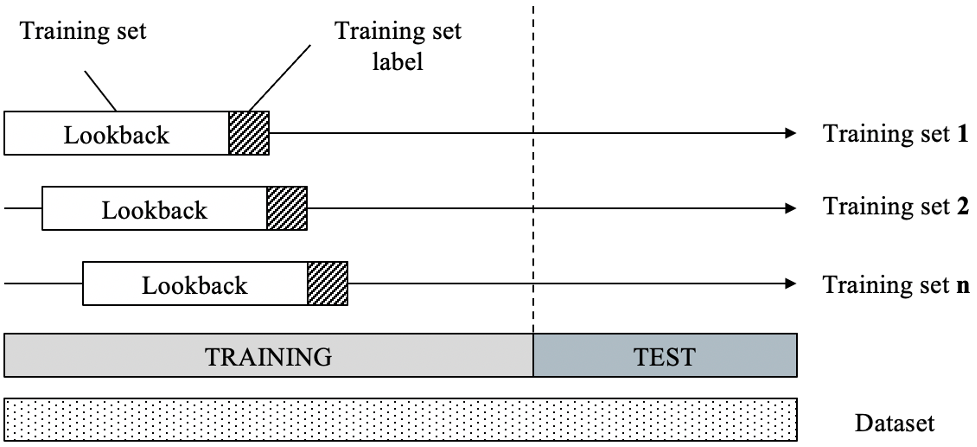}
\end{figure}

The initial dataset is divided into two parts: the training set, which is used to train the LSTM model, and the test set used to assess the model accuracy. For the latter, only the inputs are fed into the model to produce forecasts, compared with the actual values. The sliding window technique is applied to both the training and test sets.\\
\newline
A ratio of 70 and 30 percent has been used to generate training and test sets for the oil and cotton datasets. Such a choice is motivated by the amount of data that is available in the initial database. One might not select too many data points for training purposes and fewer data points for testing to avoid two main known issues: first, an overtrained model that could potentially yield overfitting and poor generalization for new inputs. Second, the lack of enough testing data points to assess model accuracy fairly. The lookback parameter is defined during the fine-tuning process with a baseline model. For a time-series with $m_{train}$ training data points and a lookback parameter $\mu$, the number of training examples is equal to $m_{train}-\mu$ for a one-unit-at-a-time ($\tau=1$) sliding window. Besides, sequential modeling with LSTM requires a 3D shape for input data. Therefore, training and test datasets are reshaped to ($m_{train}-\mu$, $\tau$, $\mu$) and ($m_{test}-\mu$, $\tau$, $\mu$) respectively.

\section{Computational aspects}
\subsection{LSTM baseline model}

To answer the research question of which of the two LSTM and ARIMA yield the best results in time-series forecasts with the selected oil and cotton datasets, they were subject to a fine-tuning process to select the best model in the model space. For LSTM, one hidden layer with 170 units baseline model was built with random hyperparameters. A min-max normalization procedure was used to accelerate the optimization process through the input data ranging between 0 and 1. The Adaptative Moment Estimation (ADAM) optimizer (Kingma and Ba, 2015) was used with a learning rate of 0.001 and an exponential decay rate of first and second moment estimates $\beta_{1}$ and $\beta_{2}$  of 0.9 and 0.999, respectively. The root mean squared error was used as a loss function, and the random generator seed was fixed to 3 for reproducibility. The baseline model was built with the Keras package on Python Anaconda's distribution. The model was executed on the Google Collaboratory Pro plan with Tensor Process Units (TPU).

\begin{table}
 \caption{LSTM hyperparameters for the baseline model}
  \centering
  \begin{tabular}{ll}
    \toprule
    \cmidrule(r){1-2}
    Hyper-parameter  & Value \\
    \midrule
    Dropout  & 10   \\
    Units      & 50 \\
    Epochs     & 50 \\
    Batch size & 32 \\
    \bottomrule
  \end{tabular}
  \label{tab:table}
\end{table}

Figures 3 and 4 shows the baseline model results for the oil and cotton datasets. Each of the figures is composed of three subplots: (i) the goodness of fit between the training dataset and its corresponding actual values, (ii) the loss values during the training process with the number of epochs, (iii) and the comparison between predicted values on the test set and its corresponding actual values. The test set predictions show that the hyper-parameters are not the optimal ones, which suggests the need for a fine-tuning process in identifying a local minimum to the cost function with its corresponding hyper-parameters values.

\begin{figure}[htbp!]
  \caption{LSTM Baseline Model results on oil datasets. (top-left): Goodness of fit on training dataset; (top-right): loss values during the training process; (bottom): Comparison between model predictions and actual values on the test set.}
  \centering
    \includegraphics[width=0.94\textwidth]{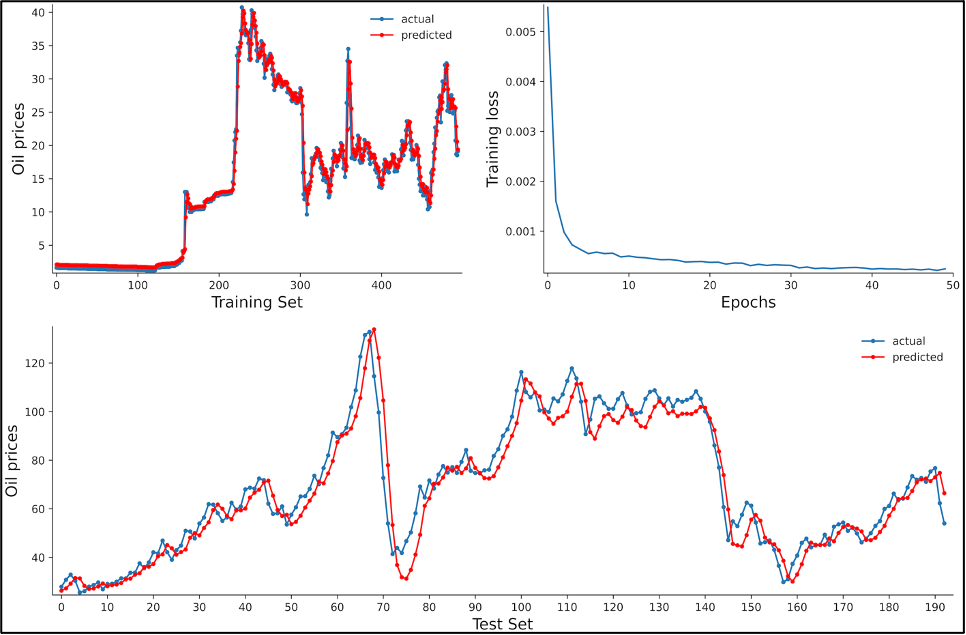}
\end{figure}

\begin{figure}[htbp!]
  \caption{LSTM Baseline Model results on cotton datasets. (top-left): Goodness of fit on training dataset; (top-right): loss values during the training process; (bottom): Comparison between model predictions and actual values on the test set.}
  \centering
    \includegraphics[width=0.94\textwidth]{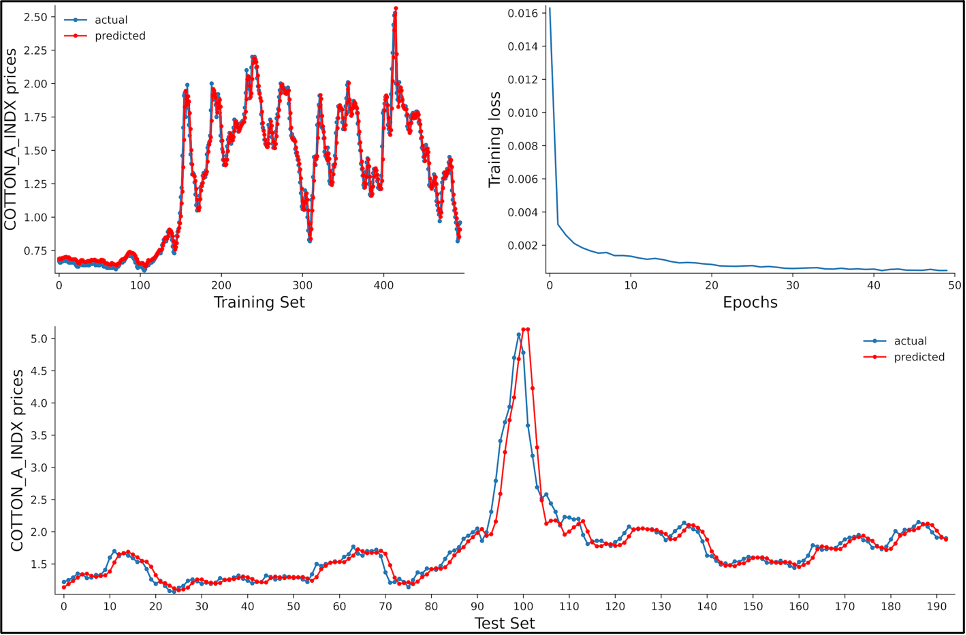}
\end{figure}

\subsection{Grid Search for LSTM Hyper-parameters tuning}

For a fair comparison between LSTM and ARIMA models, a fine-tuning process through a grid search was performed to identify a local minimum in the cost function values and its corresponding hyperparameters. Such a process is essential to identify the best model's parameter values that yield the best predictions measure with the Root Mean Square Error in our case. Four hyperparameters were considered: lookback, Dropout, LSTM units, and the number of epochs. Five values were tested and crossed via a loop to observe the best combination among them (cf. table 2).

\begin{table}
 \caption{Hyperparameters values for the LSTM grid search. Note: Lookback units are in months as the dataset}
  \centering
  \begin{tabular}{ll}
    \toprule
    \cmidrule(r){1-2}
    Hyper-parameter  & Values \\
    \midrule
    Dropout    & 0.001, 0.01, 0.03, 0.1, 0.3   \\
    LSTM Units & 10, 50, 90, 130, 170 \\
    Epoch      & 20, 40, 60, 80, 100 \\
    Lookback   & 2, 4, 6, 8, 10 \\
    \bottomrule
  \end{tabular}
  \label{tab:table}
\end{table}

An increasing number of hidden layers – starting from one - has been tested alone and yields no improvements in the error reduction, as the batch size. Similar to the baseline model, the grid search was conducted on Google Collaboratory with a TPU configuration. The process took four hours on a MacBook Pro 2019, 1.4 GHz Quad-Core Intel Core i5, 8 Gb 2133 MHz LPDDR3. The results are shown in Table 3.

\begin{table}
 \caption{Hyperparameters values that produce a local minimum for the test set RMSE for Oil and Cotton datasets}
  \centering
  \begin{tabular}{llllll}
    \toprule
    \cmidrule(r){1-6}
    Commodity  & Lookback & Dropout & LSTM Units & Epochs & RMSE \\
    \midrule
    Oil & 2 & 0.001 & 170 &	100 & 0.1974 \\
    Cotton & 2 & 0.3 & 170 & 100 & 0.1973 \\
    \bottomrule
  \end{tabular}
  \label{tab:table}
\end{table}

\section{Results}

The hyperparameters from table 3 have been used for the LSTM model and the two commodities' dataset. The results are shown in Figure 5 and 6. For both commodities, the grid search outcomes yield a faster decrease of the training loss, and the test set predictions to following better the actual data. The over and underestimations of predictions that could be observed with the baseline model and large variabilities show a better correspondence.
However, even if the grid search suggests 100 as the best number of epochs among the values provided, increasing its value does not seems to improve the learning process given the plateau reached after ten iterations. Such is aligned with observations that have been made by Sima et al. (2018).

\begin{figure}[htbp!]
  \caption{LSTM Model results for oil with hyper-parameters values retrieved from the grid search. (top-left): Goodness of fit on training dataset; (top-right): loss values during the training process; (bottom): Comparison between model predictions and actual values on the test set.}
  \centering
    \includegraphics[width=0.7\textwidth]{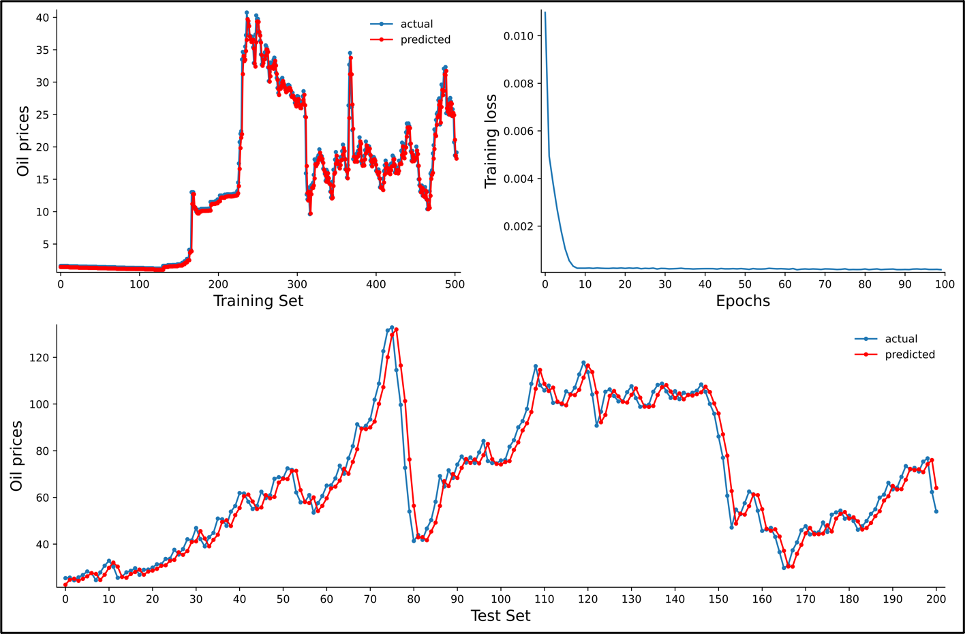}
\end{figure}

\begin{figure}[htbp!]
  \caption{LSTM Model results for cotton with hyper-parameters values retrieved from the grid search. (top-left): Goodness of fit on training dataset; (top-right): loss values during the training process; (bottom): Comparison between model predictions and actual values on the test set.}
  \centering
    \includegraphics[width=0.7\textwidth]{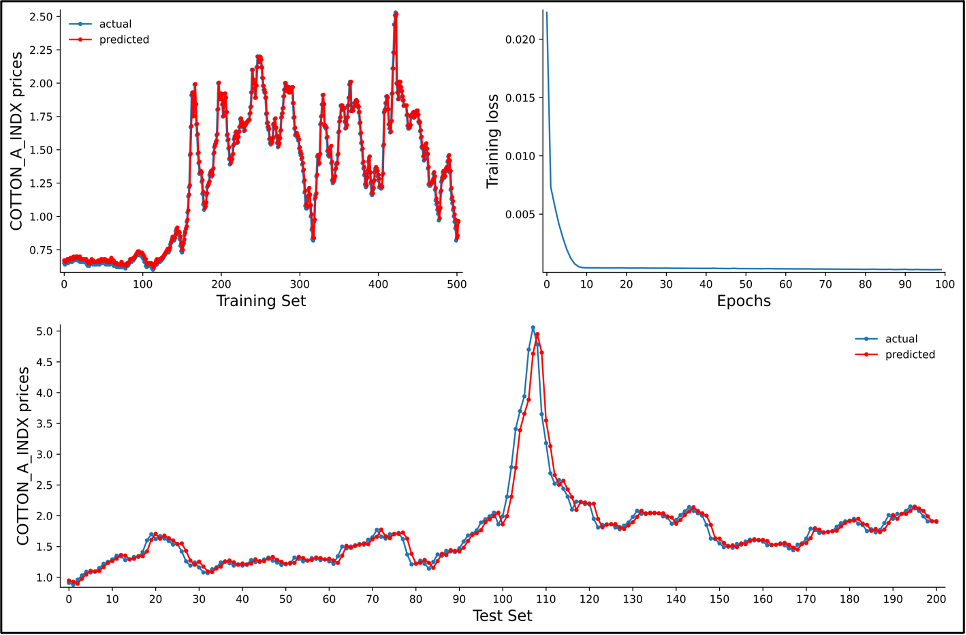}
\end{figure}

\section{Comparison with ARIMA models}  

This section compares the LSTM model results to a traditional and widely used family of models, namely the ARIMA class. First, a time series process $Y_{t}$ follows an ARMA $(p,q)$ model if it can be represented as:

\begin{eqnarray}
Y_{t} - \varphi_{i}Y_{t-i}-\cdots-\varphi_{i}Y_{t-i}=\varepsilon_t-\theta_i\varepsilon_{t-i}-\cdots-\theta_i\varepsilon_{t-i} \label{eq20}
\end{eqnarray}

Where $\varphi_{i}\left(i=1,\cdots,p\right)$ and $\theta_{i}\left(i=1,\cdots,q\right)$ are coefficients and $\varepsilon_t$ a white noise $\left(0,\sigma_{\varepsilon}^2\right)$ process.
More compactly, it can be expressed as:

\begin{eqnarray}
\Phi\left(L\right)Y_{t}&=&\Theta\left(L\right)\varepsilon_{t} \label{eq21}
\end{eqnarray}

Where $\Phi\left(L\right)=1-\varphi_{1}L-\cdots-\varphi_{i}L^{p}$ and $\Theta\left(L\right)=1-\theta_{1}L-\cdots-\theta_{q}L^{q}$ and $L$ is the lag operator.
When the $Y_{t}$ process contains $d$ (unit) roots, the process becomes an ARIMA $(p, d, q)$ and can be represented as:

\begin{eqnarray}
\Phi\left(L\right)\left(1-L\right)^{d}Y_{t}&=&\Theta\left(L\right)\varepsilon_{t} \label{eq22}
\end{eqnarray}

The first task is to perform unit root tests to select the series’ integration order $(d)$. Following perron (1989), Zivot and Andrews (1992), we perform a generalized breakpoint unit root test. Indeed, standard tests such as Dickey-Fuller have low power and are biased towards the null if the series exhibits a change in mean and or in trend. Subsequent literature has emerged, proposing various unit root tests that remain valid in the presence of a break.  The idea consists of treating each date as a potential breakpoint, and the break date, which minimizes the t-statistic associated with the Dickey-Fuller test, is selected. More formally, following Vogelsang and Perron (1998), the following generalized Dickey-Fuller test is considered with the appropriate restrictions made for each series:

\begin{eqnarray}
P_{t}&=&\mu+\beta t+\theta DU_{t}\left(T_{b}\right)+\gamma DT_{t}\left(T_{b}\right)+\omega D_{t}\left(T_{b}\right)+\alpha P_{t-1}+ \sum_{i=1}^{p}\tau_{i}\Delta P_{t-1} +u_{t} \label{eq23}
\end{eqnarray}

Where,

$P_{t}$ : represents the price of the commodity\\
$T_{b}$ : the break date\\
$DU_{t}\left(T_{b}\right)=1\left(t\geq T_{b}\right)$ : an intercept break variable which takes the value 0 for periods before the break and one after\\
$DT_{t}\left(T_{b}\right)=1\left(t\geq T_{b}\right)\bullet\left(t-T_{b}+1\right)$ : a trend break variable that takes the value 0 for periods before the break and a break date rebased trend for subsequent dates\\
$D_{t}\left(T_b\right)=1\left(t=T_{b}\right)$ :  a one-time dummy variable that takes the value one at the exact break date and 0 otherwise.\\
\newline
Special cases are obtained through restrictions on the parameters $\beta$, $\theta$, $\gamma$, $\omega$. Thus, the following restrictions yield the specific models:\\
\newline
$\beta=0$ and $\beta=0$: non-trending data with intercept \\
$\gamma=0$: trending data with intercept break\\
$\theta=0$ and $\omega=0$: trending data with trend break.\\
\newline
When no restriction is applied, we have a model with both trend and intercept breaks. Table 4 below present the results of the test allowing for breaks. For both prices, we reject the unit root hypothesis at the usual 5\% level\footnote{We also tested for seasonality using a HEGY test and no seasonal pattern was detected in the series.}.  We can then estimate an ARMA model without differencing the series. To select the best model, we analyze the series’ correlogram up to 30 lags for potential values of p and q and choose the model that minimizes the Schwarz information criterion. We ended up with an ARMA (4,2) model for cotton price and an ARMA (4,1) model for the oil price. \\
\newline
The out of sample forecasts are presented in Fig 7 and 8 and the underlying statistics in Table 5.  For both commodities, the ARIMA models perform well in predicting the prices. Table 5 shows that both the Root Mean Squared Errors (RMSE) and the Mean Average Percentage Errors (MAPE) are lower than the LSTM model.  The MAPE is 0.44 percentage point lower for cotton and 0.52-point lower for oil. To formally test the LSTM forecasts' performance vis-a-vis the ARIMA model, we perform the mean squared error-based test of Harvey-Leybourne and Newbold (1997). We conclude at the 5\% significance level that the LSTM forecasts have lower prediction errors than ARIMA for both commodities.   

\begin{table}[htpb!]
 \caption{Unit root tests. Source: Authors computation}
  \centering
  \begin{tabular}{lll}
    \toprule
    \cmidrule(r){1-3}
      & Min $t$ & $p$-value  \\
    \midrule
    Cotton & -4.94 & 0.0391 \\
    Oil & -4.85 & 0.0275 \\
    \bottomrule
  \end{tabular}
  \label{tab:table}
\end{table}

\begin{figure}[htbp!]
  \centering
  \begin{minipage}[b]{0.4\textwidth}
    \includegraphics[width=\textwidth]{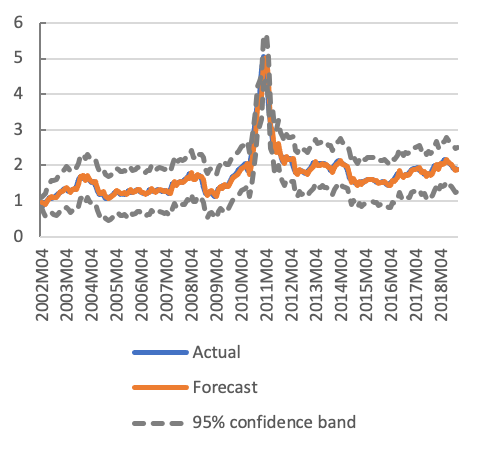}
    \caption{Cotton actual versus predicted prices. Source: Authors computations}
  \end{minipage}
  \hfill
  \begin{minipage}[b]{0.4\textwidth}
    \includegraphics[width=\textwidth]{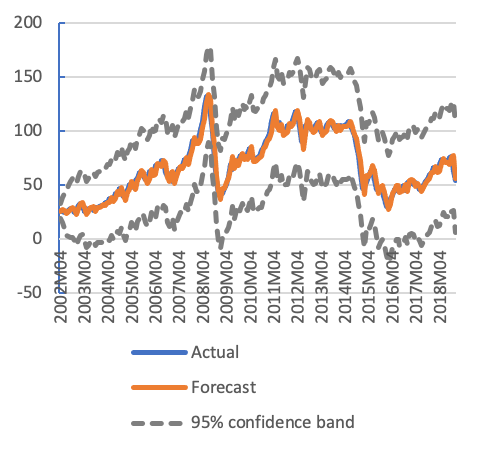}
    \caption{Oil actual versus predicted prices. Source: Authors computations}
  \end{minipage}
\end{figure}

\begin{table}[htpb!]
 \caption{Comparison of LSTM and ARIMA results}
  \centering
  \begin{tabular}{lllll}
    \toprule
    \cmidrule(r){1-5}
    & Criterion &	ARIMA &	LSTM & HLN test(p-val) \\
    \midrule
    Cotton & RMSE & 0.117 &	0.150 &	0.007*** \\
           & MAPE &	3.779 &	4.21 & \\
    \midrule
    Oil	& RMSE & 5.133 & 5.950 & 0.001*** \\
	    & MAPE & 6.432 & 6.957 & \\	
    \bottomrule
  \end{tabular}
  \label{tab:table}
\end{table}

\section{Forecast Averaging} 

There has been a recent interest in forecast averaging technics (Hendry and Clements, 2004), based on Bates and Granger (1969). The main idea behind forecast averaging (or combination of forecasts) is that models are misspecified to capture all the particular variable patterns. Indeed, a combination of forecasts can yield lower prediction errors than either of the original forecasts (Bates and Granger, 1969), since one model may make a different assumption about the relationship between the variables. \\
\newline
One question that arises when doing forecasts combination is the weights used for each model (Smith and Wallis, 2009). Different methods have been proposed in the literature, and we focus here on four of them. 

\begin{itemize}
\item Simple mean: the simple arithmetic mean of the forecasts is used for each observation. Thus, every forecast is given the same weight.\\
\item Least squares: under this approach, the forecasts are regressed against the actual values, and the coefficient of these regressions are used weights.\\
\item Mean square error weights: The Mean square error of each forecast is used to form individual forecast weights.\\
\item Mean Square error ranks: The Mean square error of each forecast is used to rank them, and then the ratio of the inverse of the ranks is used so that each forecast's weight is its rank divided by the sum of all ranks. 
\end{itemize}

Table 6 below presents the forecast combination results using different weighting schemes, and table 7 compare the best averaging model with both ARIMA and LSTM. The least-squares' weights give the best combination for the RMSE and by ranks for the MAPE for both cotton and oil. There is, however, a difference between the two commodities: while for oil, the ARIMA specification still yields the best forecasts. For cotton, the forecast averaging approach outperforms both the LSTM and the ARIMA models.  The latter is true; whatever criterion is used (RMSE or MAPE). This highlights the fact that the individual forecasts suffer from misspecification and fail to capture the variable's behavior fully.

\begin{table}[htpb!]
 \caption{Forecast combinations}
  \centering
  \begin{tabular}{lllll}
    \toprule
    \cmidrule(r){2-5}
    & Cotton & & Oil & \\
    \midrule
    & RMSE & MAPE & RMSE & MAPE \\ 
    \midrule
    Simple mean & 0.149 & 3.505 & 5.013 & 6.059 \\
    Least-squares &	0.137*** & 3.520 & 4.913*** & 6.008 \\
Mean square error &	0.145 &	3.442 &	4.962 &	6.026 \\
\midrule
Mean square error ranks &	0.143 &	3.424*** &	4.915 &	6.000*** \\
    \bottomrule
  \end{tabular}
  \label{tab:table}
\end{table}

\begin{table}[htpb!]
 \caption{Comparison of forecast averages with individual models}
  \centering
  \begin{tabular}{lllll}
    \toprule
    \cmidrule(r){2-5}
    & & ARIMA & LSTM & Average \\
    \midrule
    Cotton	& RMSE & 0.1377 & 0.1754 & 0.1374 \\ 
         	& MAPE & 3.492 & 3.935 & 3.424 \\ 
    \midrule
    Oil	& RMSE & 4.823 & 5.493 & 4.915 \\ 
        & MAPE & 5.937 & 6.366 & 6.000 \\ 
    \bottomrule
  \end{tabular}
  \label{tab:table}
\end{table}

\section{Conclusion}

Significant progress has been made over the last two decades in machine learning with potential uses in many areas, including economics. In this paper, we highlighted the potential of these new tools as an alternative or at least a complement to traditional forecasting methods. Neural network models do not require a battery of pre-tests to study the time series's underlying properties (such as unit root tests) and do not assume a parametric form.  These are two desirable properties in time series analysis. \\
\newline
While our two examples indicate that neural network models do not outperform traditional ARIMA models, the results show that combining the two approaches can yield better forecasts than either one of the models. We propose extending our analysis to a wide range of commodity prices or performing a Monte Carlo analysis to identify the situations under which one of the models will outperform the other and the circumstances for favoring a combination approach. 

\bibliographystyle{unsrt}  


\begin{thebibliography}{27}
\bibitem{Antonoviz90}Antonovitz, F. and R. Green (1990), "Alternative Estimates of Fed Beef Supply Response to risk", American Journal of Agricultural Economics, 72, 475-488.

\bibitem{Bates69}Bates, J. M and C. W. J. Granger (1969), "The Combination of Forecasts", Operational Research Quarterly, 20 (4), 451-468.

\bibitem{Box70}Box, G. E. P and G. M. Jenkins (1970), Time Series Analysis: Forecasting and Control, San Francisco: Holden-Day. 

\bibitem{Chakraborty92}Chakraborty, K., K. Mehrotra, C. K. Mohan and S. Ranka (1992), "Forecasting the Behavior of Multivariate Time Series Using Neural Networks", Neural Networks, 5, 961-970.

\bibitem{Cho14}Cho, K., Van Merriënboer, B., Bahdanau, D., and Bengio, Y. (2014, 09). On the Properties of Neural Machine Translation: Encoder-Decoder Approaches. Eight Workshop on Syntax, Semantics and Structure in Statistical Translation (SSST-8).

\bibitem{Chung14}Chung, J., Gucehre, C., Cho, K., and Bengio, Y. (2014). Empirical evaluation of gated recurrent neural networks on sequence modeling. NIPS 2014 Workshop on Deep Learning. 

\bibitem{Janvry95}De Janvry, A. and E. Sadoulet (1995), Quantitative Development Policy Analysis, Baltimore and London: The Johns Hopkins University Press.

\bibitem{FIsher35}Fisher, R. A. (1935). The Design of Experiments. Oliver and Boyd, Edinburgh.

\bibitem{Gheyas}Gheyas, I. A. and L. S. Smith, "A novel neural network ensemble architecture for time series forecasting", Neurocomputing, 74(18), 3855-3864.

\bibitem{Hendry04}Hendry, D.F. and M.P. Clements (2004), "Pooling of Forecasts", Econometrics Journal, 7, 1-31.

\bibitem{Hochreiter90}Hochreiter, S. and J. Schmidhuber (1997), "Long short-term memory", Neural Computation, 9 (8), 1735-1780. 
Kamijo, K. and T. Tanigawa (1990), "Stock price pattern recognition: a recurrent neural network approach", In Proceedings of the international joint conference on neural networks, 211-215.

\bibitem{Kingma15}Kingma, D. P., and Ba, J. (2015). Adam: A Method for Stochastic Optimization. 3rd International Conference for Learning Representations. San Diego.

\bibitem{Khandelwal}Khandelwal, I., R. Adhikari and G. Verma (2015), "Time series forecasting using hybrid ARIMA and ANN models based on dwt decomposition", Procedia Computer Science, 48, 173–179.

\bibitem{Klein50}Klein, L. (1950), Economic Fluctuations in the United States, 1921-1941, Cowles Commission Monograph No 11, New York: Wiley.

\bibitem{Kohzadi96}Kohzadi, N., M. S. Boyd, B. Kermanshahi and L. Kaastra (1996), "A comparison of artificial neural network and time series models for forecasting commodity prices", Neurocomputing, 10(2), 169-181. 

\bibitem{Kumar18}Kumar, J., R. Goomer, and A. K. Singh (2018), "Long short-term memory recurrent neural network (LSTM-RNN) based workload forecasting model for cloud datacenters", Procedia Computer Science, 125, 676–682.

\bibitem{Harvey97}Harvey, A., S. Leybourne and P. Newbold (1997), "Testing the equality of prediction mean squared errors", International Journal of Forecasting, 13, 281-291. 

\bibitem{Pai05}Pai, P.-F. and C.-S. Lin (2005), "A hybrid Arima and support vector machines model in stock price forecasting", Omega, 33(6), 497–505.

\bibitem{Perron89}Perron, P. (1989), "The Great Crash, the Oil Price Shock and the Unit Root Hypothesis", Econometrica, 57(6), 1361-1401.

\bibitem{Rumelhart86}Rumelhart, D. E., Hinton, G. E., and William, R. J. (1986). Learning representations by back-propagating errors. Nature, 323(6088), 533-536.

\bibitem{Pitman38}Pitman, E. (1938) "Significance tests which may be applied to samples from any populations: III", Biometrika, 29, 322-335.

\bibitem{Siami18}Siami Namini, Sima and Tavakoli, Neda and Siami Namin, Akbar. (2018). A Comparison of ARIMA and LSTM in Forecasting Time Series. 1394-1401. 10.1109/ICMLA.2018.00227.

\bibitem{Smith09}Smith, J. and K. Wallis (2009), "A simple explanation of the Forecast Combination Puzzle", Oxford Bulletin of Economics and Statistics, 71 (3), 331-355.  

\bibitem{Tinbergen39}Tinbergen, J. (1939), Statistical Testing of Business-Cycle Theories, Geneva: League of Nations.  

\bibitem{Vogelsang98}Vogelsang, T. and P. Perron (1998), "Additional Tests for a Unit Root Allowing for a Break in the Trend Function at an Unknown Time", International Economic Review, 39(4), 1073-1100.

\bibitem{Zivot92}Zivot, E. and D. W. K. Andrews (1992), "Further Evidence on the Great Crash, the Oil Price Shock and the Unit Root Hypothesis", Journal of Business and Economic Statistics, 10, 251-270.

\bibitem{Zou07}Zou, H. F., G. Xia, F.T. Yang and H.Y. Wang (2007), "An investigation and comparison of artificial neural networks and time series models for Chinese food grain prices", Neurocomputing,
70, 2913–2923.

\end{thebibliography}

\end{document}